\documentclass[10pt, conference, letterpaper]{IEEEtran}
\ifCLASSINFOpdf
  \usepackage[pdftex]{graphicx}
\else
\fi
\usepackage[letterspace=70]{microtype}
\usepackage{multicol}
\usepackage{enumitem}

\usepackage{array}
\newcolumntype{x}[1]{>{\centering\let\newline\\\arraybackslash\hspace{0pt}}p{#1}}

\usepackage{algorithm}
\usepackage{algorithmic}
\ifCLASSOPTIONcompsoc
 \usepackage[caption=false,font=normalsize,labelfont=sf,textfont=sf]{subfig}
\else
 \usepackage[caption=false,font=footnotesize]{subfig}
\fi
\usepackage{fixltx2e}

\usepackage{stfloats}
\begin{document}
\IEEEoverridecommandlockouts
%
\title{FilteredWeb: A Framework for the Automated Search-Based Discovery of Blocked URLs}


\author{\IEEEauthorblockN{Alexander Darer\IEEEauthorrefmark{1},
Oliver Farnan\IEEEauthorrefmark{1} and
Joss Wright\IEEEauthorrefmark{2}
}
\IEEEauthorblockA{\IEEEauthorrefmark{1}Department of Computer Science, University of Oxford, Oxford, UK}
\IEEEauthorblockA{\IEEEauthorrefmark{2}Oxford Internet Institute, University of Oxford, Oxford, UK}
}

\markboth{Network Traffic Measurement and Analysis Conference 2017}%
{}
%

\IEEEpubid{\copyright~2017 IEEE}

\IEEEspecialpapernotice{(Preprint Version - Final to appear in Network Traffic Measurement and Analysis Conference 2017)}

\IEEEtitleabstractindextext{%
\begin{abstract}
Various methods have been proposed for creating and maintaining lists of potentially filtered URLs to allow for measurement of ongoing internet censorship around the world. Whilst testing a known resource for evidence of filtering can be relatively simple, given appropriate vantage points, discovering previously unknown filtered web resources remains an open challenge.

We present a new framework for automating the process of discovering filtered resources through the use of adaptive queries to well-known search engines. Our system applies information retrieval algorithms to isolate characteristic linguistic patterns in known filtered web pages; these are then used as the basis for web search queries. The results of these queries are then checked for evidence of filtering, and newly discovered filtered resources are fed back into the system to detect further filtered content.

Our implementation of this framework, applied to China as a case study, shows that this approach is demonstrably effective at detecting significant numbers of previously unknown filtered web pages, making a significant contribution to the ongoing detection of internet filtering as it develops.

Our tool is currently deployed and has been used to discover 1355 domains that are poisoned within China as of Feb 2017---30 times more than are contained in the most widely-used public filter list. Of these, 759 are outside of the Alexa Top 1000 domains list, demonstrating the capability of this framework to find more obscure filtered content. Further, our initial analysis of filtered URLs, and the search terms that were used to discover them, gives further insight into the nature of the content currently being blocked in China.
\end{abstract}

}

\maketitle

\IEEEdisplaynontitleabstractindextext

%
\IEEEpeerreviewmaketitle

\section{Introduction}
\IEEEPARstart{F}iltering of web resources around the world is an ever changing and complex, technical and socio political activity. The aims of censorship policies depend on the region and access to the larger Internet. Advocates of free speech and expression maintain that blocking of content is detrimental to society and culture whereas censors believe this is a necessary part of controlling a large population. In response, there have been numerous attempts to discover blacklisted keywords and URLs for countries of interest~\cite{crandall2007conceptdoppler}\cite{fu2013assessing}\cite{knockel2011three}\cite{Sfakianakis2011a}. These approaches vary in the techniques used and the way they can be used to maintain these lists of filtered content. Existing approaches are, howevre, limited in the number of filtered URLs they can discover and maintain due to a lack of scalability.

We present FilteredWeb, a novel framework that allows researchers and organisations to maintain lists of filtered URLs in an automated fashion without relying on in-country expertise, human reporting, or language specific techniques. This approach uses existing infrastructure to find potentially censored URLs and can be extended based on the resources available.

We have implemented this approach as a tool, making use of a large search engine to find potentially censored domains, and exploiting features of the DNS blocking systems in target countries to test if candidate domains are indeed filtered.

Experiments demonstrate that our method is effective at discovering previously unidentified filtered URLs, and is thus a useful tool to develop and maintain continually updated lists of blocked URLs for testing. Furthermore, we have performed an initial analysis on the discovered blocked content to show that we can use this framework to collect data about the censorship landscape in a particular country.

Our tool is designed for ongoing usage in which URL lists are continually updated and tested to allow for changes due to dynamic site content and shifts in filtering behaviour.

\subsection{Contributions}
Our proposed framework is a general approach for the detection of filtered web content and can be implemented depending on circumstance. The implementation we have built, following the framework, is directly applicable for detecting poisoned domains within China. Our contributions are:
\begin{itemize}
  \item A general framework for detecting filtered URLs through the use of existing infrastructure and services.
  \item An implementation of our framework, using a specific set of services.
  \item A number of discovered poisoned domains and the search terms used to find them. 
  \item A metric to describe the effectiveness of a given base domain's patterns of language as a seed to discover further filtered domains. 
\end{itemize}

We now describe the key elements of our approach and its evaluation: the nature of Internet censorship, and textual analysis of known blocked web pages.

\IEEEpubidadjcol

\section{Filtering of the Internet}
There are many different approaches for filtering material on the Internet. The use of these depends on the network infrastructure, aims and policy of the censor. The following section describes three methods that are the most applicable for the detection framework described in this paper.

\subsection{IP Address filtering}
The blocking of IP addresses is a widely known and used method to restrict access to resources on the Internet. It is often employed within routing devices that will halt or redirect a connection to a blacklisted IP. Whilst simple to implement, over-blocking is a common problem since many websites exist on shared IPs. Furthermore, with the rise of content-delivery networks and distributed services, the IP address for certain internet services is not consistent and depends on location and the state of the resource.

\subsection{Domain Name Filtering}
\label{dnsspoof}
DNS-based filtering of the Internet is a well known approach used in many countries around the world. By manipulating the DNS, a censor can restrict access to sensitive content in an effective and scalable manner. Simply put, DNS queries for blacklisted sites can generate a range of responses~\cite{wachs2014feasibility}: \textit{an error}, \textit{an incorrect IP address} or \textit{an IP address to a non-existent server}. 

While open to circumvention by technically skilled users, this approach will stop a large portion internet users within a censored region from accessing blocked sites. The use of DNS filtering is known to be active in a number of different countries, such as China~\cite{lowe2007great}\cite{wright2014regional}\cite{Anonymous2014a}, Iran~\cite{aryan2013internet}, and Pakistan~\cite{nabi2013anatomy}\cite{aceto2016analyzing}, as well as many others.

\textit{DNS poisoning} specifically refers to a censor injecting responses for certain DNS queries, often without interfering with the original traffic or DNS servers~\cite{lowe2007great}\cite{dornseif2004government}\cite{levis2012collateral}\cite{farnan2016poisoning}. This is typically achieved either by a \textit{man-on-the-side} or \textit{on-path} attack, in which an actor passively watches DNS requests passing through key routing points in the network~\cite{Duan2012a}. 

To block domains, the censor will inject incorrect or malformed responses to queries for blacklisted domains. This injected response will often be received by the client before the \textit{genuine} response arrives, due to the censor responding to DNS queries as they are en route to the destination DNS server. In most cases, the injected response will be accepted, while the genuine one discarded---due to its arrival time after the injected. Furthermore, it is not uncommon to find that several injected responses are received \cite{farnan2016poisoning}.

In particular, a large percentage of DNS requests in China are known to be routed through areas in which poisoning occurs~\cite{lowe2007great}. Crucially, it is well known that DNS queries for blacklisted domains to random IP addresses located within the Chinese IP range will receive a response even if the destination is \emph{not} a DNS server. Furthermore, study by Farnan et al.~\cite{farnan2016poisoning} showed that DNS servers located in China are also themselves being poisoned. This means that a DNS query that is not itself intercepted will often still be responded to with an incorrect record.

\subsection{Keyword Filtering}
The blocking of requests or responses that contain blacklisted keywords is a more advanced form of internet filtering. From a technical perspective, this approach requires a more sophisticated infrastructure than other methods as TCP streams need to be parsed and searched for censored terms. For HTTP requests, the location of such blacklisted words often appear in the hostname, query or body of a request~\cite{verkamp2012inferring}. Deep packet inspection (DPI) is often used to perform this task since a censor cannot gain enough information about the request purely from meta-data. China is known to have many systems that perform this task on large amounts of Internet traffic that pass through several key routing points within their network~\cite{lowe2007great}.

\section{High Level Framework}
We propose a new approach for discovering filtered URLs within a target country. This technique uses large search engines to link patterns of language to URLs in order to find previously unknown filtered URLs. It would be impractical and inefficient simply to crawl large portions of the internet; as such, we propose a framework that uses existing infrastructure and services that already have a broad ``view'' of pages and documents on the Internet. Further, as documents are created or updated these changes will be reflected in these services.

Previous work~\cite{deibert2009geopolitics} has shown that censors have used content analysis as a part of their censorship infrastructure, and will block resources or traffic containing \textit{keywords} deemed to be sensitive. 

\subsection{Approach}
Our approach does not aim to derive blocked keywords, which is a separate area of study, but instead identifies terms that are likely to be shared by web pages that discuss key sensitive topics. Our underlying assumption is that pages that are blocked for discussing similar topics are themselves likely to contain similar patterns of language or key phrases

To begin the process, the system must be seeded with known blocked URLs for a target country. For each of these, we download the web pages and extract the text from them - discarding any HTML code, scripts or CSS. Then we extract descriptive tags from the body of text that will then be used as web search queries. 

\paragraph{Descriptive Tags} we use the notion of tags to characterise small pieces of text that describe larger bodies of content. This could be a short description of a list of individual words. Importantly, these are not necessarily keywords that are themselves filtered by a country, but instead linguistic patterns that are likely to be present on blocked web pages.

The specific method for deriving tags from web pages may vary. One key example, and the one employed in the example implementation given here, is \textit{term frequency -- inverse document frequency} (TF-IDF), which is discussed in greater detail in \ref{sec:tagextraction}. 

\begin{figure}[!t]
\centering
\includegraphics[width=1.8in]{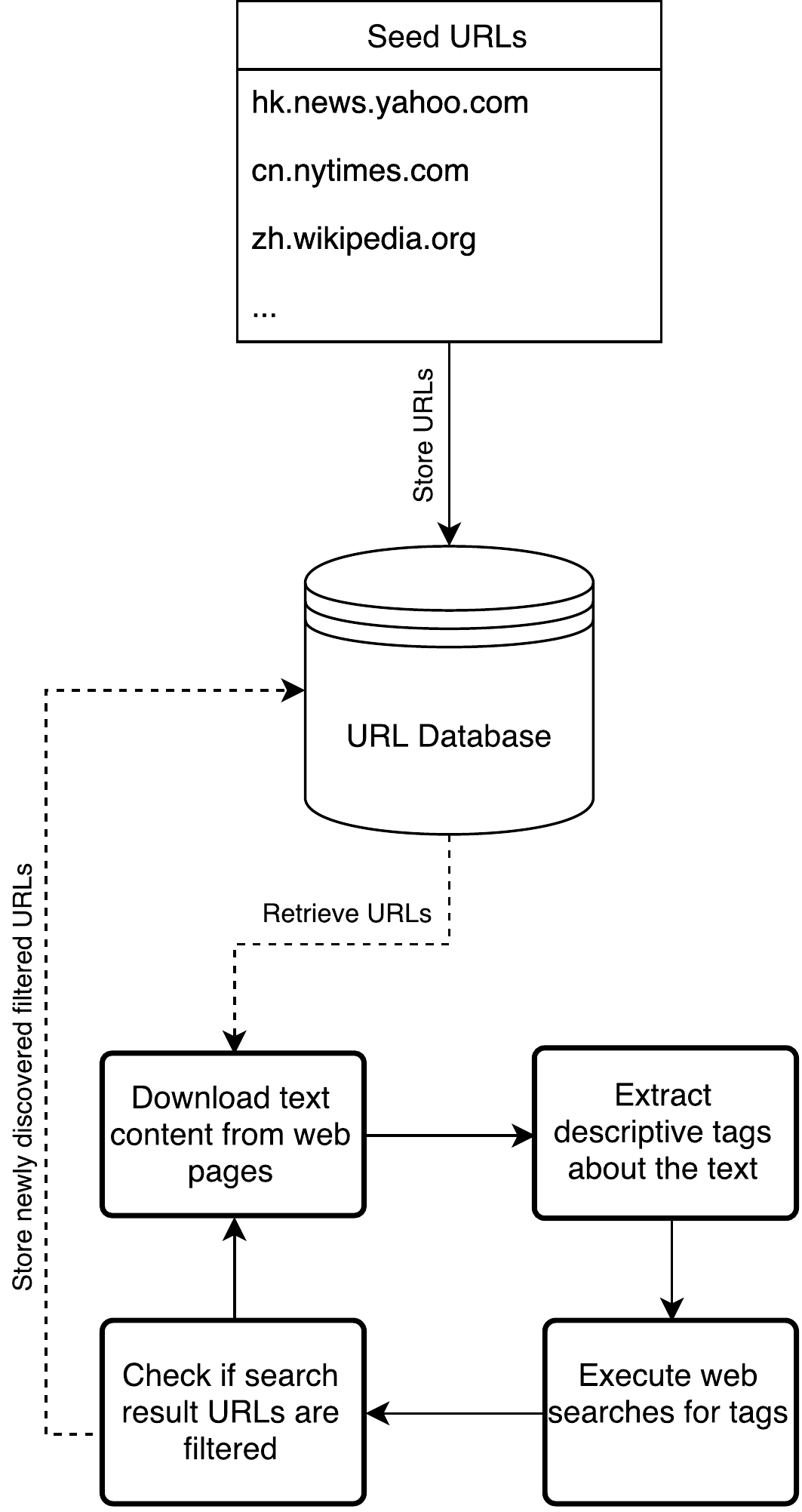}
\caption{High-level overview of the recursive filtered URL search}
\label{fig:seekflow}
\end{figure}

A web search will be made for each tag and the resulting URLs will be stored and each checked to see if they are filtered within the target country. For each newly discovered filtered URL, we will perform the tag extraction again to continue the process. This framework is summarised in Figure~\ref{fig:seekflow}. 

Our approach is recursive, and thus continually uses its results in further iterations. This results in a swift increase in the number of discovered domains, although we would eventually expect this growth to plateau. 

Further, given the dynamic nature of both web content and the behaviour of censors, any implemented approach should continually re-check filtered URLs for changes. This is something that is especially important for news sites. Pseudocode detailing the approach is shown in Algorithm~\ref{fig:seekflowalg}.

The intuition behind this approach is that filtered web pages may contain patterns of language that can then be used to discover new filtered URLs. Importantly, this framework does not identify filtered keywords, but instead uses key terms (tags) that are likely to be shared between sites discussing similar topics.

It is important to note that this framework is modular and can be cleanly separated into distinct parts:
\begin{itemize}
\item \textbf{Tag extraction:} the method by which descriptive tags are derived from a body of text.
\item \textbf{Web search:} the search engine(s) that are used to find URLs that are related to descriptive tags.
\item \textbf{Filtering check:} the method for checking if a given URL is filtered in a certain country.
\end{itemize} 
Given this, we can test for filtering of URLs in different countries by different means depending on our requirements. Furthermore, there are various existing tools and platforms that provide functionality to test if certain web resources are blocked in different countries. These include GreatFire.org\cite{greatfire}, OONI~\cite{Filasto2012a} and ICLab~\cite{iclab} among others.

Our framework is scalable since it uses openly-available and cost effective search engine APIs in order to find new URLs. This means that we do not need to build or maintain large amounts of infrastructure to crawl the Internet because this requirement is fulfilled by a search engine. Further, the more specific we can be with our search queries, the more accurate and targeted results we will achieve.

To test the effectiveness of this approach, we implement this framework in a system that will use DNS as a means for checking if URLs are filtered. This is described in Section~\ref{sec:impl}.


\begin{algorithm}
\scriptsize
\caption{Pseudocode sample for the framework}
\label{fig:seekflowalg}
\begin{algorithmic}
\STATE $urlDB \Leftarrow new~Database(seed\_urls)$
\STATE $tagDB \Leftarrow new~Database()$
\STATE $isFiltered \Leftarrow function(url)$
\STATE $getTagsFromWebPage \Leftarrow function(url)$
\STATE $doWebSearchForTag \Leftarrow function(tag)$
\LOOP
  \FOR{$url$ IN $urlDB$}
    \IF{$isFiltered(url) = TRUE$} \STATE {
      $tags \Leftarrow getTagsFromWebPage(url)$\\
      ADD $tags$ TO $tagDB$\\
    } \ENDIF
  \ENDFOR
  \FOR{$tag$ IN $tagDB$} \STATE {
      $urls \Leftarrow doWebSearchForTag(tag)$\\
      ADD $urls$ TO $urlDB$
  }
  \ENDFOR
\ENDLOOP
\end{algorithmic}
\end{algorithm}

\subsection{Ethical Considerations}
Censorship research is sensitive, and touches on a variety of ethical concerns that we must consider.

Wright et al. \cite{Wright2011a} provides an early look at the legal and ethical concerns of mapping censorship events. They deal primarily with mapping where filtering is occurring and less on the content of what is being filtered; focusing on techniques that rely on citizen volunteers to discover and report filtered content. Two later papers both published in 2015 broaden the scope of this area. The first of these by Jones et al. \cite{Jones2015a} identifies three different approaches to measuring censorship: deploying researchers with software, deploying citizens with software, and co-opting existing software. The second by Crandall et al. \cite{Crandall2015a} further differentiates censorship detection between direct observation and inference using side channel experiments. All three of these papers argue that it is important to understand the technology and motivations behind filtering, while aiming to reduce the ethical risk to individuals during the data gathering phase. 

Our approach is performed on pre-existing data. Instead of collecting data ourselves using individuals or censors, our approach uses data that has already been collected by search engines. This technique is non-invasive and---crucially---not focused on data gathered directly by researchers or citizen volunteers; thus removing the risk to individuals. While Google or Microsoft have their own set of ethical considerations---both when initially collecting the data and when making it available to others---ours become more subtly nuanced. Rather than look at the individual, we must consider the ethics of performing academic analyses of deliberately closed systems, and of using publicly available data gathered for a specific purpose to be used for different and arguably political purposes.

The first consideration is that the censorship we are observing has been installed by nation states deliberately as closed systems. They generally do not publish the details of how such systems work, or what content they are filtering. Authorities may argue that these systems are in place for the benefit of their citizens, and that obscuring their details is important to their function, or just remain silent on the subject. Is it ethical for us to observe---and publish---data about such systems?  This is considered by the aforementioned Wright \cite{Wright2011a} but he argues that these practices are of insight to us as researchers; not just to help with understanding the internet, but also to provide insight into social and political issues. Crandall et al. \cite{Crandall2015a} agree, stating that research such as this provides empirical data on censorship around the world, and that this data may be of use to political scientists and sociologists, and even to the general public.

The second consideration is of using data for purposes outside its original intent. Search companies gather data---ostensibly---to provide search results, and---cynically---to gather data on users and direct them to monetised content. Is it ethical for us to use this data to infer the details of the aforementioned closed censorship systems? Our approach relies upon the cooperation of search engines, and access through their API.  We believe that as long as their terms of use are obeyed, and our approach does not pose any risk to individuals, the ethical concerns of data used for alternative means is minimal.

\section{Implementation}\label{sec:impl}
To determine the effectiveness, efficiency, and validity of our approach, we implement the framework into a tool that can be used to find URLs that are actively filtered in China. This system is now in active use and we have used it to build a list of blocked domains that exceeds any currently published list in terms of length by several orders of magnitude. The main components for the tool are described here.

\subsection{Tag Extraction}
\label{sec:tagextraction}
To isolate descriptive tags, the documents downloaded from filtered web pages are first cleaned of any non-readable parts including HTML code, javascript and image/binary data. The remaining text is then tokenized, and each word weighted using TF-IDF~\cite{blei2003latent}\cite{rajaraman2012mining}\cite{salton1986introduction}. This results in a ranked list of words that best characterise the content of those pages in contrast to typical text in that language. At this time we only consider words that contain letters from the ISO basic Latin alphabet.

\subsubsection{Term Frequency - Inverse Document Frequency (TF-IDF)} is a statistical method of determining how important a given word is within a document. It uses a corpus (for the given language) to offset the frequency of a word as it appears in the document. Essentially, the weighting for a particular term becomes larger proportionally with the number of times it appears in the document, offset by its frequency in the corpus. Taking the \texttt{n} terms with the highest weightings can give us a list of significant keywords that characterize the content of the document.

According to best practice we remove the 1000 most common English words using the google-1000-english list~\cite{top1000}. The aim of this is reduce the number of tags that are too common and would yield generic web search results although in future implementations we can alter this to achieve a better result. We take the top 5 tags to use for web search queries.

\subsection{Web Searches}
We used the Bing search engine that is exposed through the Azure Cognitive Services API \cite{bing_search} for web search to conduct queries for the descriptive tags. This is because we use the sorting and relevance algorithms that search engine companies implement to find candidate URLs for a filtering check. Web search is a large and complicated business; most engines do not simply rank pages based on hyperlinks, but rather current trends and activity. 

One alternative to Bing is Common Crawl -- an open data project that scrapes the web for pages. While this can give us more control over the search process, the project does not provide methods for processing, sorting or querying the data. Hence why we aim to leverage the power of existing engines in order to use the ranking algorithms to find relevant links.

Another option is Baidu, however, it is known that results from this engine can be filtered and this will reduce the number of filtered URLs we can find. A study by Jiang showed that Baidu tends to drive traffic to well-known, major sites and its results raise questions about it's impartiality \cite{jiang2014search}. Furthermore, Baidu will often only direct users to site with China even if a more relevant site, given the query, exists outside of the Chinese IP range.

\subsection{Filtering Check}
We use a simple method to check if a domain is poisoned by the Chinese censorship infrastructure. As explained previously, to check if a DNS query is being intercepted, one can send it to a non-existent DNS server within China. In this case we send DNS queries to the following eight IP addresses located within China -- none of which are DNS servers:
\begin{multicols}{2}
\begin{itemize}
\item 220.181.57.217
\item 223.96.100.100
\item 1.24.10.10
\item 202.143.16.100
\item 180.160.10.1
\item 180.77.100.200
\item 144.0.111.90
\item 42.101.0.1
\end{itemize}
\end{multicols}
If a response is received to a query, this is indicative of filtering activity as the response would have originated from the filtering infrastructure: the query was intercepted and a poisoned result returned. If we receive a response to a query, we therefore mark the domain as filtered. If a response is not received after a given timeout, the filtering infrastructure was not triggered, and thus the domain is not marked as filtered.

\subsection{Parameters}
To summarise, an overview of the parameters for the system; we use:
\begin{itemize}
\item The top 5 descriptive tags from each filtered web page.
\item The top 50 search result URLs from each web search.
\item A DNS poisoning check to determine if URLs are filtered.
\end{itemize}

\section{Analysis of Approach}
We initialised our tool with URLs taken from the Citizen Lab's URL test list~\cite{czlab_lists} for China. Out of the 204 URLs present in the list, we find that 44 of the domains are DNS poisoned as of Feb 2017 and these were used to seed the system.

\subsection{Results}
Over 54,000 web searches we found 1355 poisoned domains with 115,337 filtered URLs in China. In total, our system crawled 1,113,653 unique URLs and 329,575 domains. The spread of domains to filtered URLs varies greatly and is discussed in Section~\ref{sec:further} and shown in Figure~\ref{fig:top_10_dom}. In short, 95\% of all the filtered URLs found were from just 15 (large) domains.

Table~\ref{table:doms} depicts the number of filtered URLs and domains that were discovered and the hit-rate of the tool where we calculate the number of filtered domains discovered per 1000 URLs crawled. 

It is important to note that when counting filtered domains, that each domain and sub-domain is counted separately as they may have different DNS entries. Furthermore, we have counted all Tumblr pages as a single result due to the fact that there seems to be a blanket block on all sub-domains (Tumblr works by giving each of it's users a sub-domain for their pages).

\begin{table}[!t]
\renewcommand{\arraystretch}{1.3}
\caption{Discovered filtered URLs as of Feb 2017}
\label{table:doms}
\centering
\scriptsize
\begin{tabular}{|c||c|}
\hline
& \textbf{Counts} \\
\hline\hline
\textbf{Unique URLs Crawled} & 1,113,653 \\ \hline
\textbf{Unique Domains Crawled} & 329,575 \\ \hline
\textbf{Unique Filtered URLs} & 115,337 \\ \hline
\textbf{Unique Poisoned domains} & 1355 \\ \hline
\textbf{Filtered URLs / 1000 URLs crawled} & 103.57 \\ \hline
\textbf{Poisoned domains / 1000 domains crawled} & 4.11 \\ \hline
\end{tabular}
\end{table}

\begin{figure}[!t]
\includegraphics[width=2.9in]{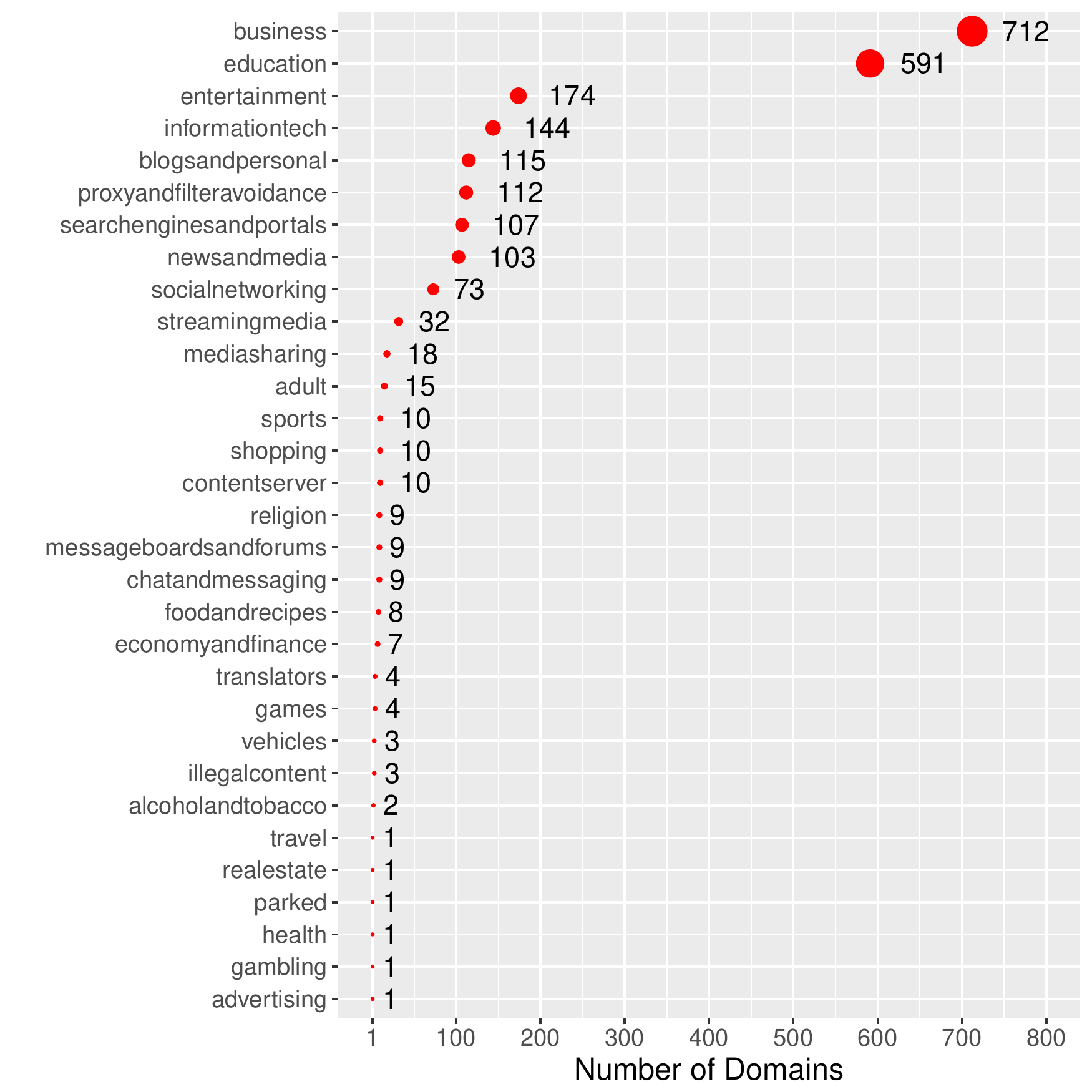}
\caption{Category Breakdown for Poisoned Domains}
\label{fig:dom_cats}
\vspace{2mm}
\centering
\emph{Note: some domains appear in more than one category}
\end{figure}

For further analysis, we categorised all of the discovered filtered domains to understand the nature of sites that are currently being blocked. This was done using the WebShrinker Categories API \cite{webshrinker}, a service that indexes URLs and classifies them into categories. These results are shown in Figure~\ref{fig:dom_cats}.

\subsection{Discussion of Results}
Our testing has demonstrated that the framework we propose is able to discover a significant number of filtered URLs that were not present in initial seed lists. These results indicate that there is an exploitable connection between the content of individual filtered web pages and other filtered domains. Further, existing search engines can demonstrably be used as tools to uncover censored material on a large, automated scale. We plan to release a list of the poisoned domains discovered; however, due to space limitations, we cannot reproduce it in this paper.

An advantage of this approach is that \textit{only} filtered URLs are used as the basis of linguistic patterns for search requests. This increases the efficiency of the tool and prevents unnecessary crawling of large portions of the Internet to discover newly-filtered URLs.


The system has discovered a large number of domains that are currently being filtered in China. The hit rate shows that the tool has to crawl about 1000 URLs to find 4 \emph{poisoned} domains. In this instance, this approach could have been optimised by limiting the crawling to domains that were not already stored in the database. For this initial test however, it was necessary to crawl larger numbers of pages in order to fully test the usefulness of the framework. In a future version, these optimisations could increase the efficiency and reduce the cost of further discovery. Furthermore, we could start to reduce the crawls of larger more well known sites - such as facebook.com and twitter.com. This would lower the total number of URLs crawled by a substantial amount, however, it is not clear if these social media services actually benefit the system by providing very new and constantly changing content -- further studies into these mechanics is a potential avenue for future research.

In comparison to lists available via the Citizen Lab -- shown in Table~\ref{table:cit}, we have found a great deal more filtered URLs. Given our tool checks for DNS poisoning, we can have a high confidence in the numbers of blocked web pages we have found. Moreover, our list is continually updated and growing.

\begin{table}[!t]
\renewcommand{\arraystretch}{1.3}
\caption{Comparison with Alternative Filtered URL lists for China}
\label{table:cit}
\centering
\scriptsize
\begin{tabular}{|x{3.5cm}||c|c|c|}
\hline
& \textbf{FilteredWeb} & \textbf{Citizen Lab} & \textbf{CensMon} \\
\hline\hline
\textbf{Filtered URLs} & 115,337 & 204 & N/A \\ \hline
\textbf{Poisoned Domains} & 1355 & 44 & 176 \\ \hline
\textbf{Filtered URLs - Top 1000 Removed} & 4153 & 166 & N/A \\ \hline
\textbf{Poisoned Domains - Top 1000 Removed} & 759 & 34 & N/A \\ \hline
\end{tabular}
\vspace{2mm}\\
\emph{The list of censored domains discovered by CensMon is not publicly available, therefore we only have the number reported.}
\end{table}

\begin{figure*}[!t]
\centering
\subfloat[Discovery of filtered domains over time (cumulative)]{\includegraphics[width=2.0in]{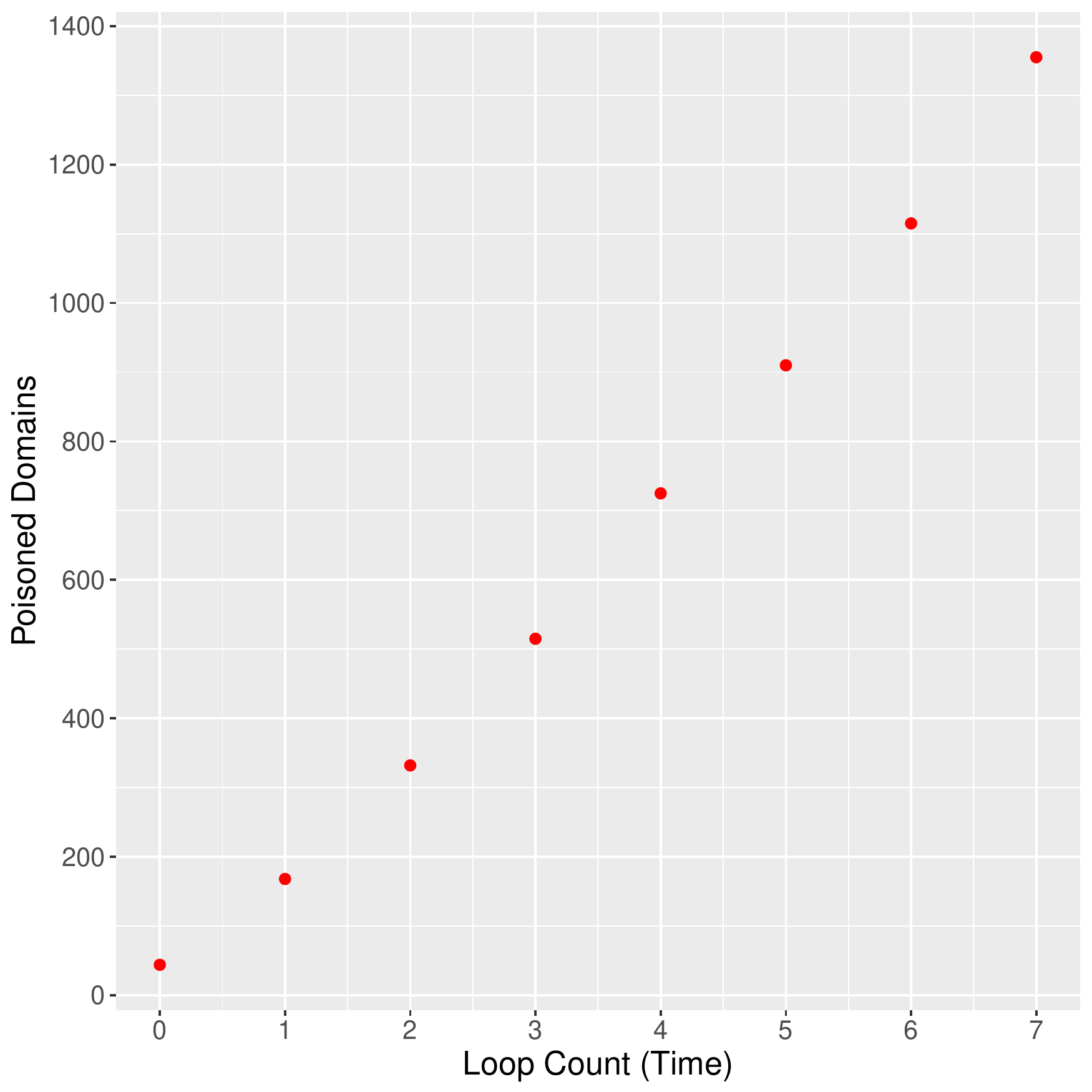}
\label{fig:dom_recur}}
\hfil
\subfloat[Discovered Poisoned Domains against Crawled URLs]{\includegraphics[width=2.0in]{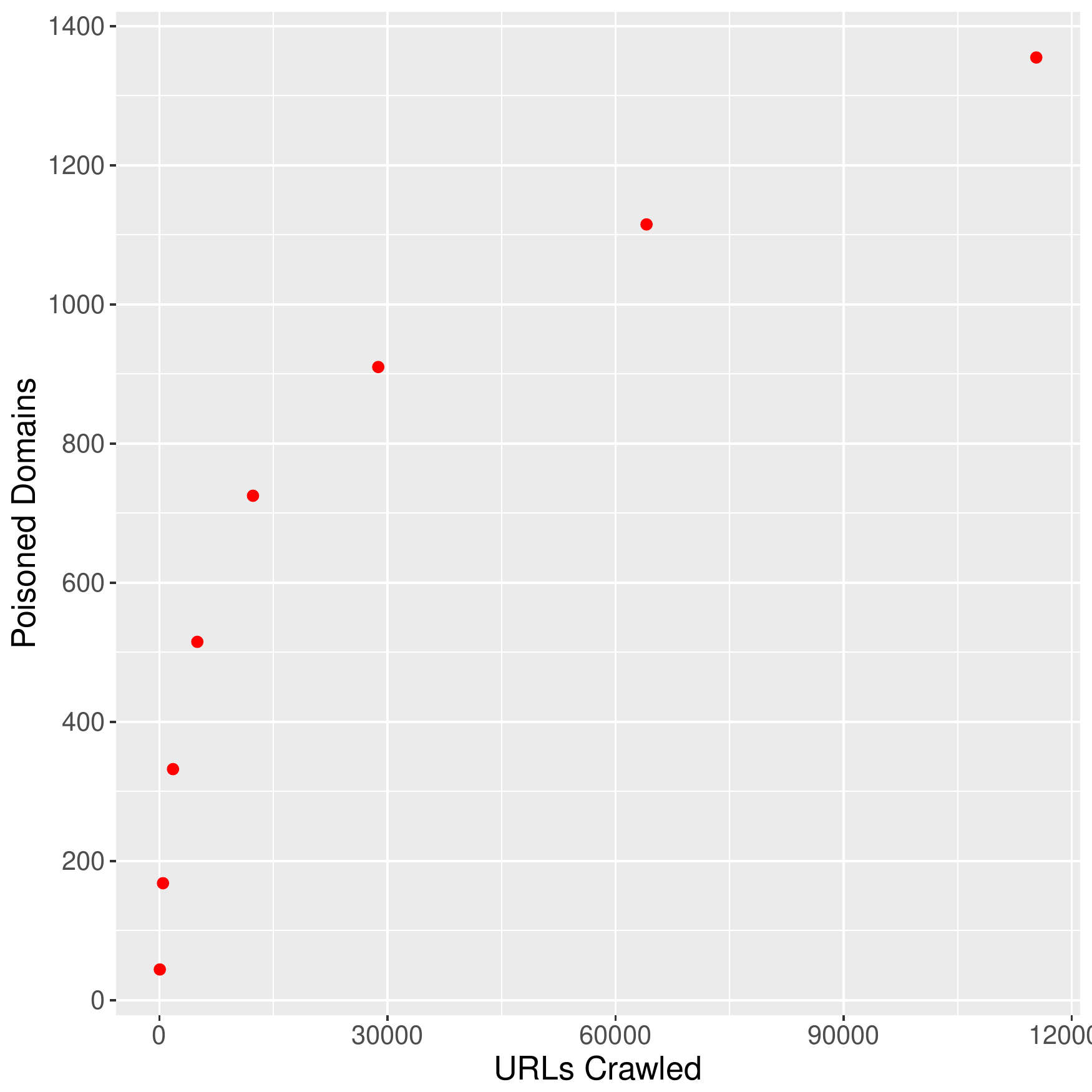}
\label{fig:dom_vs_urls}}
\caption{Tool Performance}
\label{fig_sim}
\end{figure*}

Our tool is recursive, this means we can monitor the number of filtered domains it discovers over time. This metric is useful in determining the effectiveness of the approach since we can compare the number of domains crawled against the number of filtered domains found. Figure~\ref{fig:dom_recur} shows the number of poisoned domains discovered over time, or each loop of the system; from this we can see that the discovery rate is relatively constant. Figure~\ref{fig:dom_vs_urls} shows the number of discovered poisoned domains against the number of URLs crawled.

From the categories of the poisoned domains, shown in Figure~\ref{fig:dom_cats}, we find there are a large number of business focused sites, which are domains used for the sale of services and products. Further, there seems to be a substantial effort to block domains for educational sites and services as well as entertainment, filtering avoidance and news. This appears to be in line with work published by King et al. \cite{king2013censorship} that posits that the Chinese government aims to reduce \textit{collective action potential} rather than individuals. Moreover, King et al. provide evidence showing that censors in China is not particularly harsh towards critique of the censorship policies themselves.

\subsection{Further Analysis}
\label{sec:further}
\begin{figure*}[!t]
\includegraphics[width=0.95\textwidth]{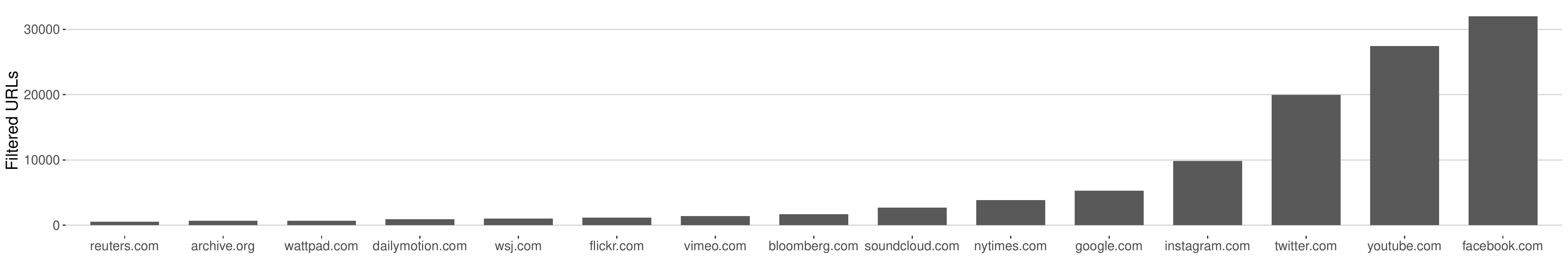}
\caption{15 Most Frequently Occurring Poisoned Domains}
\label{fig:top_10_dom}
\end{figure*}

Analysing the database of URLs collected by our tool revealed a number of interesting artifacts about its operation, and thus the state of censorship in China. Firstly, we look at the most commonly occurring domains that we crawl, from this we can see the larger and better known sites and social media services make up a large portion of the URLs we have processed. Approximately 95\% of the filtered URLs we find are from only 15 domains -- these are shown in Figure~\ref{fig:top_10_dom}.

\begin{figure*}[!t]
\includegraphics[width=0.95\textwidth]{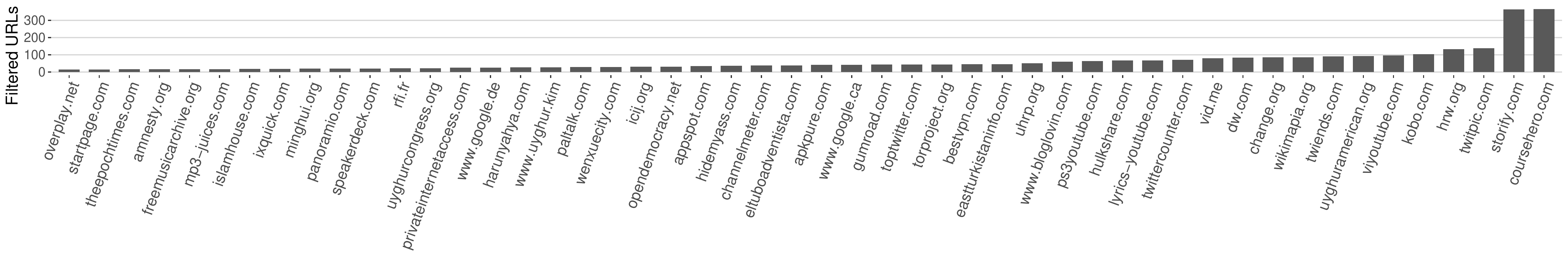}
\caption{50 Most Frequently Occurring Poisoned Domains with Alexa Top 1000 Removed}
\label{fig:top_50_dom_alexa_removed}
\end{figure*}

Due to the fact that domains within the Alexa Top 1000 are the most regularly occurring within our database, we have produced Figure~\ref{fig:top_50_dom_alexa_removed} that shows the 50 most common poisoned domains outside of the Top 1000. These domains appear far less frequently in web search results, but importantly, they can give us a deeper insight into what the Chinese government is currently blocking.

\begin{figure*}[!t]
\includegraphics[width=0.95\textwidth]{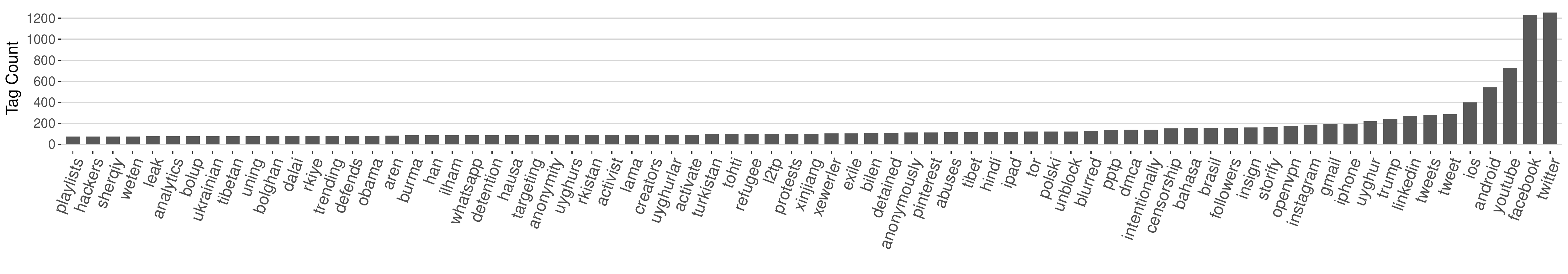}
\caption{Top 75 Descriptive Tags found in Filtered Web Pages}
\label{fig:top_100_tags}
\end{figure*}

An interesting use of the collection of descriptive tags is that we can search for them \textit{ex post facto} on each of the filtered URLs discovered. From this we can see which tags feature most prominently in poisoned domains within China. This is by no means a search for necessarily blocked keywords, however, it could be used as a method to find patterns of language or topics that may have caused a page or domain to be blocked. Figure~\ref{fig:top_100_tags} shows the top 75 descriptive tags across all the filtered URLs in our database after removed the Alexa Top 1000\footnote{We do this because we found that many of the top 1000 have similar and generic language, we are more interested in the subtler and potentially more sensitive language.}.

\subsection{Discovery Power}
The basic assumption of our approach is that filtered web pages contain linguistic patterns that could lead to further filtered URLs. Our framework achieves this by deriving descriptive tags from a base URL which are used as search queries to link to other URLs---where the tag is the connection.

To convey how strongly certain domains provide descriptive tags that lead to the discovery of other filtered material, we introduce a metric: \textit{discovery power}. This represents the number of filtered URLs from different domains that are found by searching with tags isolated from a given base domain. We calculate the discovery power by looking at the tags that, when used as search queries, result in the discovery of at least one other poisoned domain. (We do not count instances within the search results where the discovered domain is the same as the base domain---that is to say we don't count self-discovery.)

\begin{figure*}[!t]
\centering
\subfloat[Number of discovered filtered URLs against the number of descriptive tags used that resulted in the discovery of filtered URLs]{\includegraphics[width=3.4in]{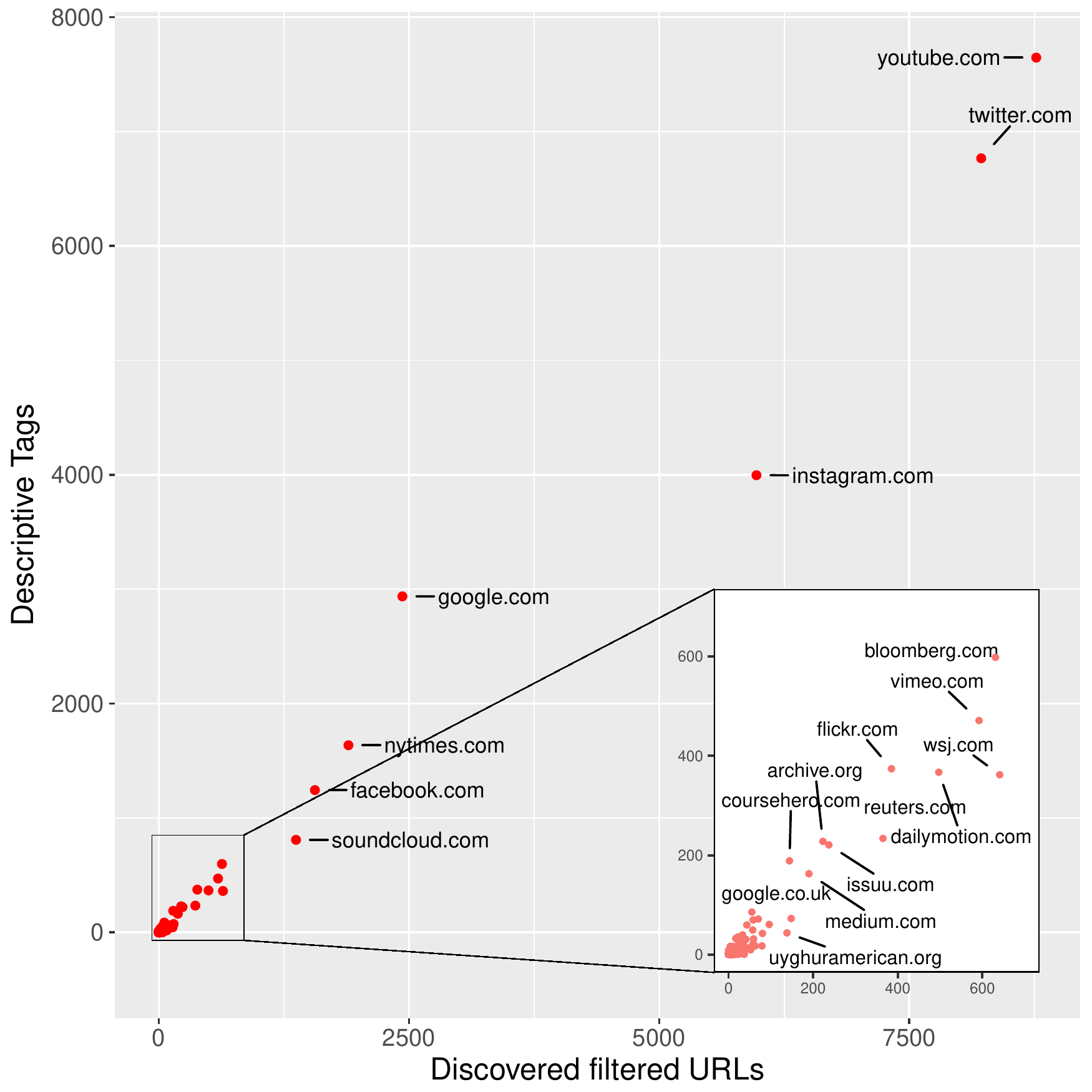}
\label{fig:dom_strength}}
\hfil
\subfloat[Extension of plot in Figure~\ref{fig:dom_strength}, with Alexa Top 1000 removed]{\includegraphics[width=3.4in]{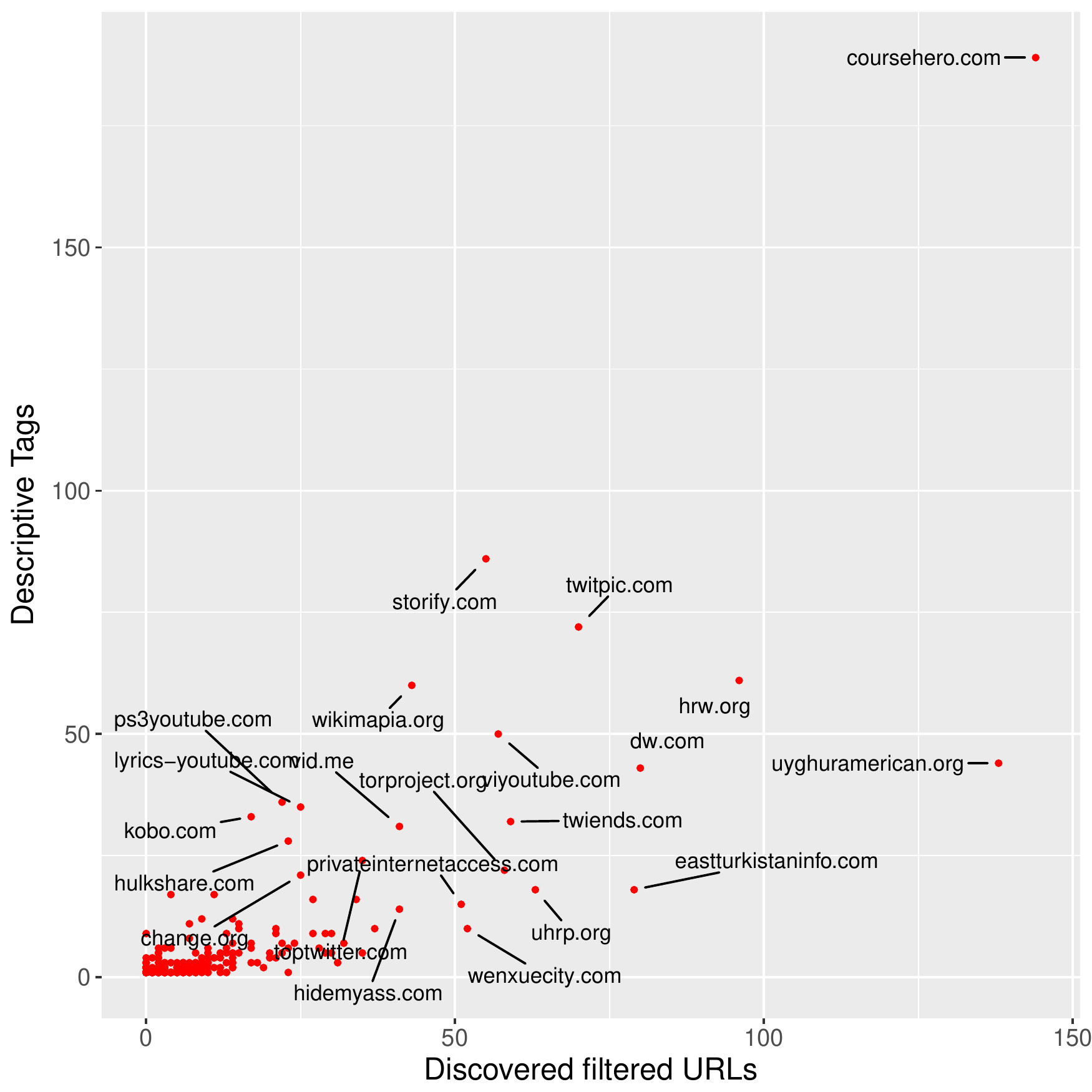}
\label{fig:dom_strength_top_1000}}
\caption{Filtered Domain Discovery}
\label{fig_sim}
\end{figure*}

Comparing the discovery power with the number of crawled URLs for each domain provides insight into how much potentially sensitive language is present on those domains. This is not only interesting in the understanding of censorship activity, but also to optimise our tool to consider only higher power domains when deriving tags. Figure~\ref{fig:dom_strength} shows how many descriptive tags were used to find filtered URLs against the number of filtered URLs found for each poisoned domain. Figure~\ref{fig:dom_strength_top_1000} shows the same, but with the the Alexa Top 1000 domains removed from the dataset. 

Notable examples in Figure~\ref{fig:dom_strength_top_1000} are: uyghuramerican.org, dw.com, hrw.org and eastturkistaninfo.com, all of which produced ``high-power'' tags that resulted in a higher number of filtered URLs than the number URLs crawled for each of those domains.

\subsection{Limitations of Implementation}
There are three main limitations we have identified with our implementation of this framework. The first is that the act of filtering may result in search engines being unlikely to list filtered web pages highly in their rankings due to a lack of other sites linking to them. Similarly, for smaller websites, it may be the case that search engines do not rank these highly enough to be included in our approach. This could be mitigated by using one or more alternative search engines and combining the results. However, even given this, we still get good results when using just a single engine.

A second limitation is the means of checking if a URL is censored in a particular country. Whilst DNS censorship is widely practiced it is difficult to know definitively if a web page is blocked through sole use of this technique. Use of alternative testing frameworks, such as OONI or ICLab, would mitigate this limitation significantly and allow for more targeted testing.

Moreover, there are other, more sophisticated techniques for detecting blocking that have been published~\cite{wright2014regional}\cite{Filasto2012a}. The use of DNS in our tool is due to its scalability and speed. The downside of this is that we will not detect if URLs are filtered by other means.

Finally, we only consider English words when isolating descriptive tags. This means that a large chunk of potentially filtered (Chinese) sites are being missed by our tool. A future version of the system will include Chinese words and phrases as well as their translation in English.

\section{Related Work}
Various pieces of research over recent years have aimed to detect filtered web resources~\cite{Aceto2015b}. A notable example is CensMon~\cite{Sfakianakis2011a} which is is an architecture that can be used as a censorship monitoring tool for the web. This system uses a combination of social media, Google Trends and user input to generate candidate URLs to test for filtering. By taking a search-based approach, and feeding our results recursively back into our dataset, our novel use of linguistic patterns as search queries to link URLs has yielded a list of blocked domains several orders of magnitude larger than CensMon's reported results.

The use of DNS as a means for detecting censorship is used in a study by Wander et al.~\cite{wander2014measurement}. They show that a major advantage of DNS is the fact we can perform the analysis from outside the target countries by injecting queries into the networks within them. The authors used this approach to test for filtered domains within China and Iran.
Other country specific studies conducted for China include~\cite{lowe2007great}\cite{wright2014regional} and for Iran~\cite{aryan2013internet}\cite{anderson2013dimming}.

While our work has focused on discovering filtered URLs, it is closely related to studies that aim to derive filtered keywords or phrases. ConceptDoppler, a system proposed by Crandall et al.~\cite{crandall2007conceptdoppler}, attempts to do this for China using \textit{latent semantic analysis} on known keyword block lists to discover semantically related terms that may also be blocked. While similar, ConceptDoppler is orthogonal to the framework presented in this paper, since we are focused the discovery and testing of filtered URLs rather than the keywords themselves, which is the motivation and focus of ConceptDoppler.

There have also been a number of user-based systems proposed for detecting Internet filtering, notably UBICA~\cite{Aceto2015a} and  work by Winters~\cite{Winter2013a}. These are dependent on volunteers to gain insight into local networks by either providing information or running automated probes from within the target countries. These are fundamentally different approaches to ours since we conduct all of our analysis from a single location outside the target networks.

\section{Conclusions}
We have presented a framework for the automated discovery of filtered websites without reliance on contextual knowledge or language specific information. The approach is fully automated, scalable and effective. Our experiments have demonstrated that the approach discovers previously unknown filtered content, which yields further meaningful reults when fed back into itself. The system does not require complex infrastructure---beyond pre-existing public services---or significant computational or network requirements to operate, but instead uses existing public and cost-effective resources.

Further, this approach does not rely on local knowledge or participation from individuals in countries of interest, and thus avoids many significant ethical challenges in building filter lists. Our technique is limited by the effectiveness of the keyword mining methods and censorship checking methods. While these present challenges, our results already demonstrate significant results with real world application.

The results we have generated are a significant contribution. Our list of currently poisoned domains within China is far more up to date and accurate than what is currently available and we expect to make this available in the near future. We expect this list to change and grow over time. With the framework we have proposed, this can be done with minimal effort and resources. Even with our early results, we present a filter list of poisoned domains for China that is 30 times larger than the current most widely-used public list.

Finally, this paper has presented an early analysis using the data collected through the use of this framework. The ability to find URLs that can lead to other blocked material is substantial and has considerably increased the capability for us to monitor and understand internet censorship.

\section{Future Work}
The current limitations of the implementation lend themselves to easy extension. The most obvious, and important, of these is to extend the target of the tool to apply to other states that implement network filtering. Clearly, this requires integration with methods that allow automated checking of filtering in other countries beyond China, however, we have proposed a number of means to do this and are actively pursuing this potential.

More directly, the tool should be expanded to incorporate multiple search engines for a more thorough and unbiased view of the Internet. 

The extraction of longer key phrases from filtered URLs as search terms should be investigated as a means to provide more specific search results. This could include the use of n-grams or other, more sophisticated language processing techniques. There are a number of existing services that can provide functionality for creating short descriptions from larger bodies of text, or indeed web pages. These models are usually machine learning based so could be custom designed for this purpose. Even so, some existing services are able to leverage large amounts of data that has been scraped from the Internet.

Finally, incorporating a range of methods for checking the filtered status of resources, and the use of existing testing infrastructures such as OONI or ICLab, would greatly increase the ability of the tool to build comprehensive filtering lists.

Outside, the extensions to the approach, there is significant scope for additional and more thorough analysis of the data we have collected. Our database of URLs, descriptive tags and poisoned domains lends itself to further study and research, both in the topic of filtered content and in trends in filtering over time. Additionally, measuring the discovery power of base domains could give significant insight when trying to discover patterns of censorship. The work presented here presents a good starting point for that venture.

\bibliographystyle{IEEEtran}

\end{document}